# Collective Motion of Inelastic Particles between Two Oscillating Walls

Fei Fang Chung, Sy-Sang Liaw, and Wei Chun Chang

*Department of Physics, National Chung-Hsing University*
*250 Guo-Kuang Road, Taichung 402, Taiwan*

**Abstract**

This study theoretically considers the motion of *N* identical inelastic particles between two oscillating walls. The particles' average energy increases abruptly at certain critical filling fractions, wherein the system changes into a solid-like phase with particles clustered in their compact form. Molecular dynamics simulations of the system show that the critical filling fraction is a decreasing function of vibration amplitude independent of vibration frequency, which is consistent with previous experimental results. This study considers the entire group of particles as a giant pseudo-particle with an effective size and an effective coefficient of restitution. The *N*-particles system is then analytically treated as a one-particle problem. The critical filling fraction's dependence on vibration amplitude can be explained as a necessary condition for a stable resonant solution. The fluctuation to the system's mean flow energy is also studied to show the relation between the granular temperature and the system phase.

# I. Introduction

Granular systems lose energy due to inter-particle collisions. Mechanical agitation is usually used as an energy supply to keep a granular system active. This kind of excited granular system has already attracted much attention due to its many attractive phenomena, including pattern formation [1,2], segregation [3-5], and phase transition [6-12] which are commonly observed in experiments. The kinetic theory [13] and hydrodynamics model [14,15] are commonly used for statistical analysis of the various behaviors of granular systems. On the other hand, the system as a whole can't be well understood without careful study of the dynamics of individual particles. Therefore, bouncing balls on vibrating tables remains an active research focus [16-18].

Liquid-solid transitions [6] were discovered in the vertical vibrated granular systems, and were observed in recent experiments on a horizontal vibrating system [19-21] consisting of a monolayer of spherical particles rolling on a plate. The plate is driven horizontally by external harmonic force and the particles gain energy by interacting with the substrate and boundary walls, with gravity and air having a negligible effect. When the particles' filling fraction reaches a critical value, the structure and dynamics of the system sharply shift from irregular to regular. Experimental data [21] show that the critical filling fraction is inversely proportional to the vibration amplitude, a counter-intuitive outcome given that larger amplitudes, and thus greater energy, would tend to drive the system into a disordered state. This is actually an example of "Freezing by heating" which was first introduced by Helbing et al. in 2000 [22]. This phenomenon also occurs in other systems such as colloidal suspension [23] and granular segregation [20]. Here we theoretically investigate the monolayer granular system under horizontal vibration. Based on evidence from molecular dynamic (MD) simulations, this study will show that a highly energetic ordered state can be sustained in the vibrating granular system when a resonance occurs together with inelastic clustering [24-26], and the dependence of the critical filling fraction on amplitude can also be explained by the resonant condition.

This report considers $N$ identical particles of a diameter $d$ confined between two oscillating walls. The walls vibrate sinusoidally in phase in the $x$-direction. The inelastic collisions between the particles, and between the particles and the walls, are characterized by the same coefficient of restitution $\varepsilon$. Friction is neglected in the theoretical analysis and its effects are discussed in the comparison of the theoretical results with the experiments. In the case of $N=1$, the particle can move in resonance to the vibrating walls with a fractional frequency $f/n$ [27,28], where $f$ is the frequency of the vibrating walls. We solve the analytic solution of resonant velocity for $n=1$ in Sec. II. In Sec. III, we use MD simulations to study the energy and configuration of many-particle systems. We find that inelastic clustering occurs when $N$ is large and $\varepsilon$ is smaller than a critical value so that the particles move collectively as a giant particle. We use the resonant conditions for this pseudo particle to determine the critical filling fraction at any given vibrating amplitude. In Sec. IV, we analyze the velocity distribution of the particles. In a gas-like (i.e., dilute) phase, the velocity distribution is well fitted by the superposition of two Gaussian distributions, showing the coexistence of two granular temperatures [29,30]. In a solid-like (i.e., compact) phase, the velocity distribution is a superposition of one Gaussian and one exponential distribution [31]. We discuss how the critical filling fraction depends on the coefficient of restitution in Sec. V. Sec VI is the conclusion.

## II. Stable resonant solutions for a one-particle system

Let the walls oscillate horizontally with velocity $v_p = A\omega \sin\omega t$ (Fig. 1). Suppose a particle is moving towards the left with velocity $-v$, $v > 0$. When the particle collides with the wall at instant $t$, it changes its velocity to $v'$. By the law of conservation of momentum and the definition of coefficient of restitution, we have:

$$v' = \varepsilon v + (1+\varepsilon)v_p = \varepsilon v + (1+\varepsilon)A\omega\sin\omega t \tag{1}$$

where the mass of the wall is assumed to be infinite. The particle then moves with constant velocity $v'$ to the right and hits the right wall at instant $t'$. In the time interval $t'-t$, the particle travels a distance $(L-d) - A\cos\omega t' + A\cos\omega t$, where $A\cos\omega t$ and $A\cos\omega t'$ are the displacements of the left and right wall from their vibrating centers at time $t$ and $t'$, respectively. So we have:

$$v'(t'-t) = (L-d) - A\cos\omega t' + A\cos\omega t \tag{2}$$

When the motion of the particle is in resonance with the vibration, we expect $v' = v \equiv v_R$ from the symmetry of the system. The particle takes a half period of time to travel from the left wall to the right wall and vice versa. Therefore, the phase difference between two collisions at opposite walls is $\omega t' - \omega t = \pi$. In the resonant state, Eqs. (1) and (2) are, replacing variable $t$ by the resonant phase divided by the angular speed $\theta_R/\omega$:

$$v_R = \varepsilon v_R + (1+\varepsilon)A\omega\sin\theta_R \tag{3}$$
$$v_R \pi/\omega = L - d + 2A\cos\theta_R \tag{4}$$

and can be solved for $v$ and $\theta$ at resonance:

$$v_R = \frac{\pi\omega(L-d) \pm 2\omega\sqrt{(4\alpha+\pi^2)A^2 - \alpha(L-d)^2}}{4\alpha+\pi^2} \tag{5}$$

$$\theta_R = \tan^{-1}\left(\frac{2\sqrt{\alpha}v_R}{\pi v_R - \omega(L-d)}\right) \tag{6}$$

where $\alpha = \left(\frac{1-\varepsilon}{1+\varepsilon}\right)^2$. The condition for the existence of real solution $v_R$ is that the value in the square root in Eq. (5) has to be non-negative, which can be expressed as:

$$A/L \geq \sqrt{\frac{\alpha}{4\alpha+\pi^2}}(1-d/L) \tag{7}$$

The condition says that the vibration amplitude $A$ cannot be too small to have a resonant solution.

A solution for Eqs. (3) and (4) must be stable to be physically accessible. To determine the stability of the solution, we go back to Eqs. (3) and (4). A small perturbation $(dv, d\theta)$ to the solution $(v, \theta)$ will induce a change $(dv', d\theta')$ for the velocity and phase at the next collision:

$$\begin{pmatrix} dv' \\ d\theta' \end{pmatrix} = \begin{pmatrix} \varepsilon & (1+\varepsilon)A\omega\cos\theta_R \\ \dfrac{-\varepsilon\pi}{v_R - A\omega\sin\theta_R} & \dfrac{-(1+\varepsilon)\pi A\omega\cos\theta_R}{v_R - A\omega\sin\theta_R} \end{pmatrix} \begin{pmatrix} dv \\ d\theta \end{pmatrix} \equiv \mathbf{M} \begin{pmatrix} dv \\ d\theta \end{pmatrix} \tag{8}$$

The eigenvalues of the matrix $\mathbf{M}$ are $\lambda_1 = 0$ and $\lambda_2 = \varepsilon - \dfrac{\pi(1+\varepsilon)A\omega\cos\theta_R}{v_R - A\omega\sin\theta_R}$. A stable solution exists only when $\lambda_2 < 0$. Substituting the solutions for $v$ and $\theta$ given in Eqs. (5) and (6) into the expression of $\lambda_2$, we can write the inequality $\lambda_2 < 0$ in the following form:

$$A/L \geq e(\varepsilon)(1 - d/L) \tag{9}$$

where the coefficient $e(\varepsilon)$ is a function of $\varepsilon$ only, independent of the driving frequency $\omega$. One can show that $e(\varepsilon)$ is always larger than the factor $\sqrt{\dfrac{\alpha}{4\alpha + \pi^2}}$ shown on the rhs of Eq. (7), which means that producing a larger $A$ requires a stable resonance. We plot the resonant solutions, $v_R$ and $\theta_R$ as a function of $A/L$ for the system with $d/L=0.2$ and $\varepsilon=0.8$ in Fig. 2, where the stable and unstable solutions are respectively indicated by the solid and broken lines.

## III. N-particle systems

For many ($N$)-particle systems we used molecular dynamics (MD) (see Appendix) to calculate the x-component velocity of the center of mass (CM) $v_{CM} = \dfrac{1}{N}\sum_{i=1}^{N} v_{xi}$ and determine the impact phase $\theta_{CM}$ between the group and the walls by monitoring the instant when $v_{CM} - v_p$ changes its sign. For all the simulations reported in this paper, we fixed the frequency at $f = 1/2\pi$. We ran the simulation for 200 vibration periods and used the particle trajectories in the last 100 periods to calculate the time average impact phase $\langle \theta_{CM} \rangle$ and the mean flow energy $\langle E_{CM} \rangle = \dfrac{1}{100T}\int_{100T}^{200T} v_{CM}^2 dt$, where $T = 1/f$ is the period of driving force.

We first simulate the one-particle ($N=1$) case and compare the results with the analytic results of the last section. Fig. 2 shows the simulation results of the average speed and the impact phase as a function of relative amplitude $A/L$, together with the analytic results given by Eqs. (5) and (6). The length of the container and the coefficient of restitution are $L=5d$ and $\varepsilon = 0.8$, respectively. According to Eq. (7), resonant solutions exist for $A/L \geq 0.0283$. However, a solution is not stable unless $A/L \geq 0.0464$ according to Eq. (9), where $e(\varepsilon = 0.8) = 0.0581$. Therefore, for an $A/L$ below 0.0464, the CM of the particles hits the two walls at random phases so that the average speed is low. Starting from $A/L = 0.0464$, the average speed jumps to a higher value, and the impact phase at each wall converges to a constant value as predicted by the theoretical stable solution.

We repeat the simulation for the two-dimensional case with $L \times W = 20d \times 10d$ at a different amplitude $A$ and a different number $N$ of particles. The results show that the system transits from a gas-like state to a solid-like state at a critical number of particles for a given amplitude. We plot $\langle E_{CM} \rangle$ and $\langle \theta_{CM} \rangle$ as a function of the filling fraction $\phi = N\pi d^2 / 4LW$ at $A$=2.5$d$, 3$d$ and 3.5$d$ in Fig. 3. Two distinct regions separated at a critical filling fraction $\phi_c$. For $\phi < \phi_c$, $\langle E_{CM} \rangle$ decreases rapidly when the filling fraction increases. $\langle \theta_{CM} \rangle$ is random and thus yields an average value with a large fluctuation. For $\phi \geq \phi_c$, $\langle E_{CM} \rangle$ increases to a local maximum and then decreases gradually to the average energy $A^2\omega^2/2$ of the external drive and $\langle \theta_{CM} \rangle$ converges to an average value with a small fluctuation similar to the resonant case of the one particle system. The differences of these two states can be further revealed from the configuration of the particles and the motion of the center of mass. For this, we calculate the x-component position of the center of mass $x_{CM}(t)$ as a function of time to study the mean flow motion of the system and the pair correlation function [*] $g(r)$

$$g(r) = \frac{2LW}{aN(N-1)} \sum_{i=1}^{N} \sum_{j=1}^{i-1} H(r_{ij} - r)H(r + \Delta - r_{ij}) \qquad (10)$$

to describe the structure of the particles. The term $a = \pi(2r + \Delta)\Delta$ is the area of a ring with an inner radius of $r$ and a width $\Delta = 0.1$. $H(x)$ is the Heaviside function and $r_{ij}$ is the distance between particles $i$ and $j$.

In Fig. 4, we plot $\langle g(r) \rangle$ and $x_{CM}(t)$ for three different filling fractions at $A$=3$d$ with $\varepsilon = 0.8$. At $\phi = 0.079$, $x_{CM}(t)$ oscillates back and forth with irregular amplitude at the same frequency as the driving walls. The corresponding $\langle g(r) \rangle$ shows no obvious structural peaks, and we refer to it a gas-like state. At $\phi = 0.314$, the peaks of $\langle g(r) \rangle$ at $r/d$=1 and 2 appear, indicating that the number of direct contacts between the particles has increased. As a result, more energy is dissipated due to inelastic collisions, causing $x_{CM}(t)$ to fluctuate with a small amplitude near the middle of the two walls. At $\phi = 0.511 > \phi_c$, $x_{CM}(t)$ oscillates periodically with a uniform amplitude implying that the particles move collectively in resonance to the driving walls. In the mean time, another distinct peak of $\langle g(r) \rangle$ occurs at $r/d$ ~1.73, implying a hexagonal packing. We classify this state as a solid-like state. The critical filling fraction $\phi_c$ we define here indicates the transition point of the two states: the chaotic motion with loose packing and the collective motion with close packing.

In the solid-like state, the compact cluster of particles can be approximated by a giant pseudo particle moving in resonance to the external drive with a velocity $v_R$ given by Eq. (5) except that the size $d$ and the restitute coefficient $\varepsilon$ have to be replaced respectively by the effective values $d_{eff}$ and $\varepsilon_{eff}$ for the pseudo particle. We can write $d_{eff} = \delta L \phi$, where the geometric factor $\delta$ has a value of around 1. In Fig. 3, we plot $v_R^2 / A\omega^2$, with $\varepsilon_{eff} = 0$ [28], as a function of $\phi$ for a solid-like state in solid lines. This fits the simulation result of $<E_{CM}>/A^2\omega^2$ very well. The condition for the existence of real value $v_R$ now reads

$$\delta\phi \geq 1 + \sqrt{4 + \pi^2 A/L} \qquad (11)$$

The critical filling fractions $\phi_c$ obtained from the equality in Eq. (11) agree very well (Fig. 5) with the simulation results obtained from Fig. 3.

Quantitatively, our theoretical values of $\phi_c$ are lower than the experimental values [20,21]. The difference comes from our neglect of friction in the simulations between the particles and the bottom of the container which dissipates the energy of the particles. For a given $\phi$, more energy is needed from the external drive to compensate for energy loss through friction so that $E_{CM}$ can gain enough energy to meet the resonant condition. Thus, the phase transition occurs at a larger $A$. On the other hand, for a given $A$, $E_{CM}$ is smaller than the value when friction is neglected. To satisfy the resonant condition, more particles are needed to reduce the distance traveled by CM from one wall to the other. Thus phase transition occurs at a larger value of $\phi_c$.

**IV. Granular temperature**

The total energy of the particles is the sum of the mean flow energy $E_{CM}$ and the fluctuation energy defined as $\frac{1}{N}\sum_{i=1}^{N}(v_i - v_{CM})^2$ which is usually used for defining granular temperature $T_g$:

$$T_g = \frac{1}{N}\sum_{i=1}^{N} u_i^2, \quad u_i = v_i - v_{CM} \qquad (12)$$

We have seen in the previous section that $E_{CM}$ of the solid-like state in the two-dimensional case can be obtained approximately by the completely inelastic particle model and the model provides a good prediction to the critical filling fraction. Here we will see that $T_g$ also behaves differently in the gas- and solid-like states. In Fig. 6 we plot the average granular temperature $<T_g>$, scaled by $A^2\omega^2$, as a function of $\phi$. We see that $<T_g>$, like $<E_{CM}>$ as shown in Fig. 3, changes abruptly at a critical filling fraction $\phi_c$. Surprisingly, $<T_g>$ increases when the system alters its state from gas-like to solid-like, unlike ordinary fluidization [6] which has a low value for $T_g$ in the solid-like state. It has been shown [32] that for a dilute granular gas with a Maxwellian velocity distribution, the granular temperature can be determined by equating the rate of energy dissipated and the power input to yield $T_g \sim A^2\phi^{-1}$. For our system, $<T_g>$ is proportional to $A^2$ and a decreasing function of $\phi$ by a power of $-1.8\pm0.2$. We plot the velocity distribution $P(u)$ for $A=3d$ and $\varepsilon=0.8$ in Fig.7. These distribution curves can be well fitted by the superposition of two Gaussian distributions in the gas-like state [29,30] while, in the solid-like state, they are more like the superposition of an exponential and a Gaussian distribution [31]:

$$P(u) = \begin{cases} \dfrac{N_1}{N}e^{-u^2/2T_1} + \dfrac{N_2}{N}e^{-u^2/2T_2}, & \text{gas-like} \\ \dfrac{N_1}{N}e^{-|u|/\sqrt{T_1/2}} + \dfrac{N_2}{N}e^{-u^2/2T_2}, & \text{solid-like} \end{cases} \qquad (13)$$

Fig. 8 shows the fitted values for temperatures $T_1$ and $T_2$, and particle portions $N_1/N$ and $N_2/N$ as a function of $\phi$. $N_1 > N_2$ and $T_1 < T_2$, that is, a lesser portion of particles ($N_2/N$) hitting the walls has a larger temperature ($T_2$), while a larger portion of particles ($N_1/N$) which do not have direct contact with the walls has a smaller temperature ($T_1$). Because the two particle density domains have different temperatures in gas-like state, the average temperature $T_g$ as a function of $\phi$ does not simply follow the power law $\phi^{-1}$, but rather $\phi^{-1.8}$ as seen in Fig. 7. We can calculate $T_g$ analytically using the velocity distribution Eq. (13) to obtain

$$T_g = \frac{T_1 + kT_2}{1+k} \qquad (14)$$

where $k = \dfrac{N_2}{N_1}\sqrt{\dfrac{T_2}{T_1}}$ for the gas-like state and $k = \dfrac{N_2}{N_1}\sqrt{\dfrac{\pi T_2}{T_1}}$ for the solid-like state. Based on the values for $T_1$, $T_2$, $N_1/N$, and $N_2/N$ given in Fig. 8, we found the value of $k$ is small compared to 1 but $kT_2$ is always comparable with $T_1$. It turns out that the numerical value of $T_g$ is roughly two times that of $T_1$ for both the gas- and solid-like states. It would be interesting to see whether the relation $T_g \approx 2T_1$ has any explanation or is simply a special case in our system.

## V. Dependence of the restitute coefficient

Our discussion above is based on simulation results given $\varepsilon=0.8$. In Fig. 9 we show both the average total energy, $<E>$, and the pair correlation function at $r=d$, $<g(d)>$, as a function of filling fraction at $\varepsilon=0.8, 0.85, 0.87, 0.9, 0.95$, and $0.98$. When $\varepsilon \to 1$, c approaches the Carnahan-Starling equation [33]. We see that $\phi_c$ is insensitive to the exact value of $\varepsilon$. However, the gas- and solid-like states become indistinguishable when $\varepsilon$ is larger than the value $\varepsilon_c \approx 0.93$. When $\varepsilon > \varepsilon_c$, $<E>$ becomes a monotonically decreasing function without a local maximum or minimum and the discontinuity of the slope of $\langle g(r) \rangle$ disappears. That is, when $\varepsilon > \varepsilon_c$, inelastic clustering does not occur and the granular temperature would be so high that particles would be unable to form a stable cluster. The situation is similar to water condensation where steam will never condense to water beyond a critical temperature regardless of pressure. According to a simplified model called independent collision wave (ICW) approximation reported by Bernu and Mazighi [24], the value of $\varepsilon_c$ that causes $n_c$ particles in a one-dimensional system to collapse into a cluster is given by $\varepsilon_c \approx 1 - \pi / n_c$. For a two dimensional system $n_c$ is the number of particles per column width [26]. Here, we use $\phi_c$ to estimate the value of $n_c$ and then calculate $\varepsilon_c$ to obtain a value roughly 30% larger than the predicted ICW approximation, but this deviation is expected. Since our particles are confined in a two-dimensional finite region and energy is steadily injected, the interference of collision waves may not be neglected so that particles in our system cluster more readily than in the ICW approximation.

## VI. Conclusion

We simulated the motion of *N* identical inelastic particles between two oscillating walls and studied the properties of the gas- and solid-like phases of the system at different filling fractions, driving amplitudes and coefficients of restitution. The transition requires two conditions: (1) the filling fraction of the particles and the driven amplitude must satisfy the resonant condition given by Eq.(11); (2) the particle's coefficient of restitution has to be small enough that the inelastic clustering can occur at the filling fraction which satisfies the resonant condition. At the transition point, both of the mean flow energy and the fluctuation energy increase dramatically. We found that the mean flow energy of the solid-like state can be well approximated by the resonant energy of a pseudo particle in the completely inelastic model. The resonant condition gives the value of critical filling fractions which is a decreasing function of vibration amplitude, consistent with previous experimental results. We have also found, from an analysis for the fluctuation energy, that the gas-like state of our system can be described by a dilute particle system with two coexistent temperatures, one for particles gaining energy directly from the walls and the other for those gaining energy from inter-particles collisions.

## Appendix—Simulation method

We consider *N* identical spheres of diameter *d=2* and mass *m=1* which are confined between two oscillating walls. The two walls are in simple harmonic motion in direction *x* at amplitude *A* at frequency $f = 1/2\pi$. Besides the dissipative collisions which are characterized by the coefficient of restitution $\varepsilon$, there is no other way to dissipate energy from the system. Frictions and rotations of the particles are not considered in our simulation. Energy is supplied by the particles striking the walls.

The collision force $\vec{F}_{ij}$ between the *i*'th and *j*'th particles is modeled by the dissipative linear spring force:

$$\vec{F}_{ij} = -\left[\frac{m}{2}\omega_s^2\left(d - |\vec{r}_{ij}|\right) + 2\gamma(\vec{u}_{ij} \cdot \vec{r}_{ij}/|\vec{r}_{ij}|)\right]\vec{r}_{ij}/|\vec{r}_{ij}|$$

where $\vec{r}_{ij} = \vec{r}_j - \vec{r}_i$ is the vector along the line connecting the centers of particles *i* and *j*, and $\vec{u}_{ij} = \vec{u}_j - \vec{u}_i$ is the relative velocity. When the contact duration of collision $t_c$ and the coefficient of restitution $\varepsilon$ are specified, the parameters $\omega_s$ and $\gamma$ can be determined by

$$\gamma = -\frac{\ln \varepsilon}{t_c} \quad , \quad \omega_s = \sqrt{(\pi/t_c)^2 + \gamma^2}$$

We have used $t_c = 3 \times 10^{-3}$ s in all computations.

We integrate the equation of motion:

$$\frac{d\vec{v}_i}{dt} = \begin{cases} \sum_{j=1} \vec{F}_{ij} & \text{if } |\vec{r}_i - \vec{r}_j| < d_i + d_j \\ 0 & \text{else} \end{cases}$$

$$\frac{d\vec{r}_i}{dt} = \vec{v}_i$$

by Verlet's Method [34] with time step $dt = t_c/30$. Collisions between particles and walls are also treated by the spring force with the same $\varepsilon$.

**Acknowledgments**

This work is supported by the National Science Council of Taiwan under the grant number NSC97-2112-M005-005, and the National Center for Theoretical Sciences of Taiwan.

**Figure Captions**

Fig. 1

Schematic diagram of one sphere of diameter $d$ confined in a rectangular container. Two walls in $x$-direction are at a constant distance $L$ apart and in simple harmonic motion with amplitude $A$ and frequency $\omega/2\pi$.

Fig. 2

(a) Average speed and (b) impact phase at two walls of the particle moving between two oscillating walls. The parameters used in simulation are $L$=10, $\varepsilon = 0.8$, and $d = 2$. Results of MD simulations (diamond) are consistent with the analytic predictions for unstable (broken line) and stable (solid line) solutions.

Fig. 3

Time average values of mean flow energy $\langle E_{CM} \rangle$ and impact phase $\langle \theta_{CM} \rangle$ as a function of filling fraction at $A$=2.5$d$, 3$d$ and 3.5$d$. The frequency is $f = 1/2\pi$ and the coefficient of restitution is $\varepsilon = 0.8$.

Fig. 4

The pair correlation function $g(r)$ and x-component of CM $x_{cm}(t)$ for three different filling fractions $\phi =$ 0.079, 0.314, and 0.511 at $A$=3$d$ with $\varepsilon = 0.8$. The inset graphs at right corner of first column are the corresponding snapshot configurations of particles at $t$=128$T$.

Fig. 5

Dependence of the critical filling fraction on the relative amplitude for 2-dim case(filled circle), $L \times W = 20d \times 10d$. The frequency is $f = 1/2\pi$ and the coefficient of restitution is $\varepsilon = 0.8$. The solid line is the result of the equality in Eq. (11) with $\varepsilon_{eff} = 0$ and $\delta = 1.24 - 1.6A/L$. The circle symbols are the experimental results of the two-dimensional system reported in Ref. [21].

Fig.6

(a) Time average of fluctuation energy $\langle T_g \rangle$, scaled by $A^2 \omega^2$, as a function of filling fraction $\phi$ at the parameters same as figure 3. The solid lines are plotted according to Eq.(14) with parameters shown in Fig.

8.

Fig. 7

Velocity distribution function of (a) gas-like state at $\phi$=0.118, 0.157, 0.196, and 0.236, and (b) solid-like state at $\phi$=0.511, 0.589, 0.668, and 0.746. The solid lines are the fitting curves given by Eq.(13) with the parameters values shown in Fig. 8.

Fig.8

The temperatures $T_1$, $T_2$ and particle portions $N_1/N$, $N_2/N$ respectively as a function of the filling fraction. With these values, Eq. (13) fits the velocity distribution (Fig. 7) approximately.

Fig. 9

(a)The pair correlation function at contact $\langle g(d) \rangle$ and (b) the time average energy $\langle E \rangle / A^2\omega^2$ as a function of filling fraction at $\varepsilon$=0.8(+), 0.85(□),0.87(×), 0.9(○), 0.95(▲), and 0.98(◇). The dash line in (a) is the Carnahan-Starling equation for nearly inelastic disk.

Fig.1

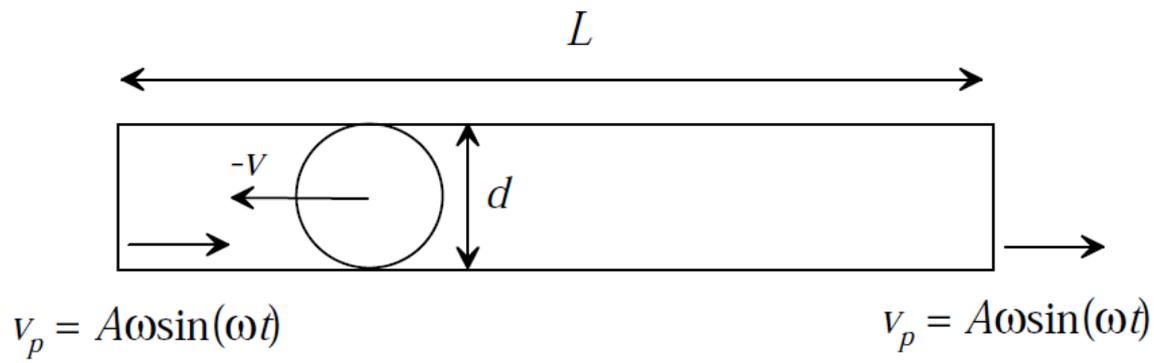

Fig. 2

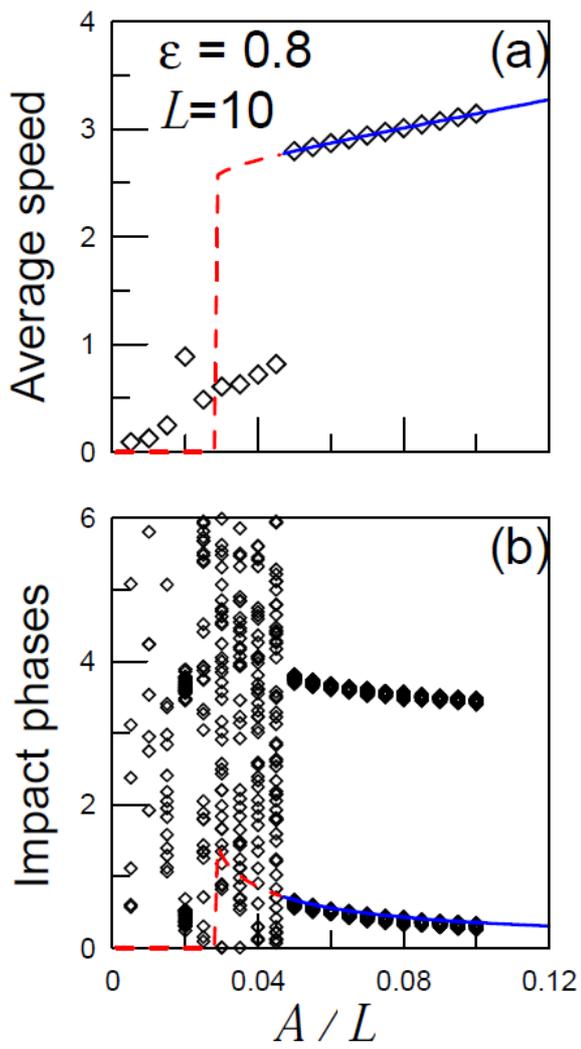

Fig. 3

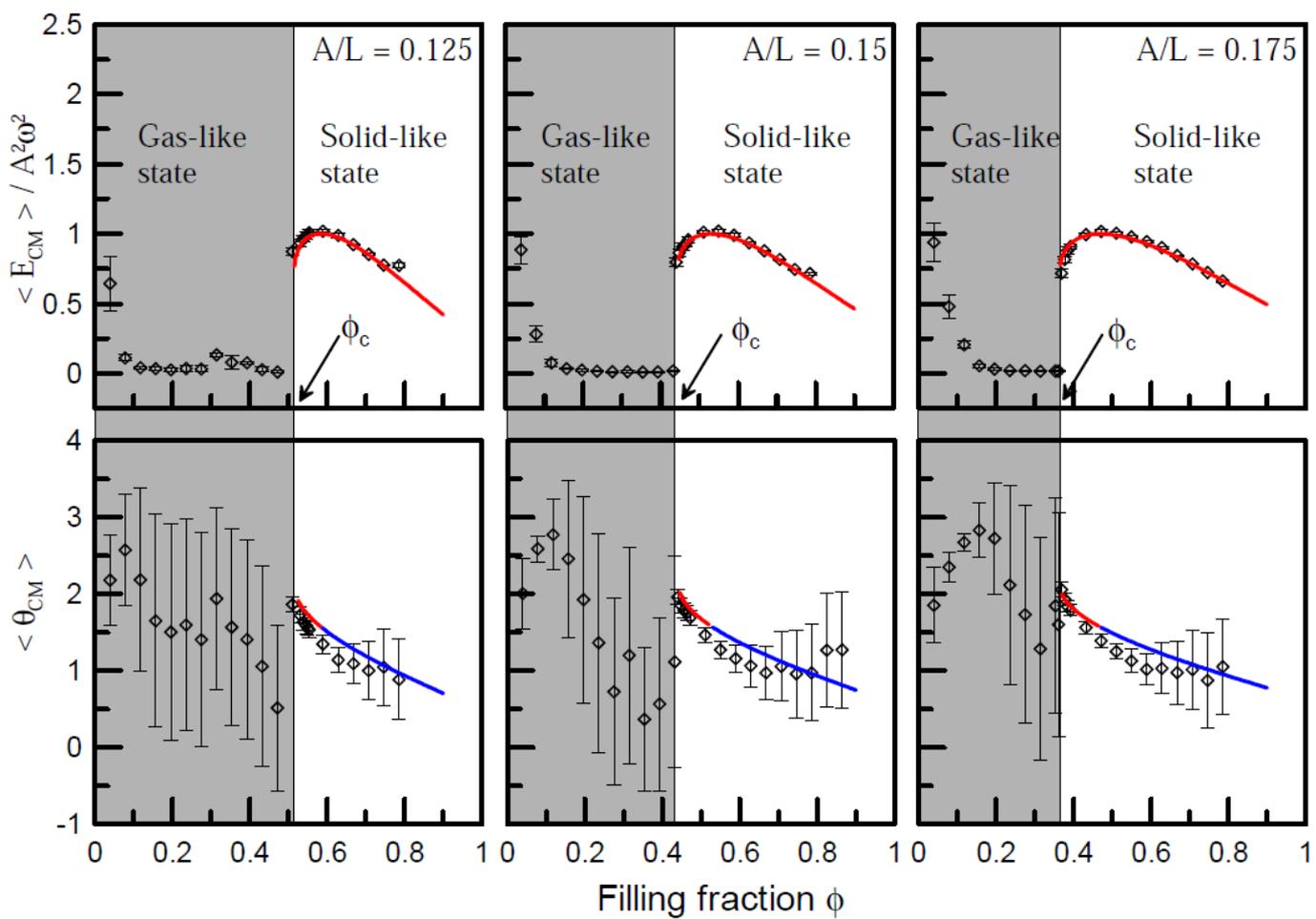

Fig. 4

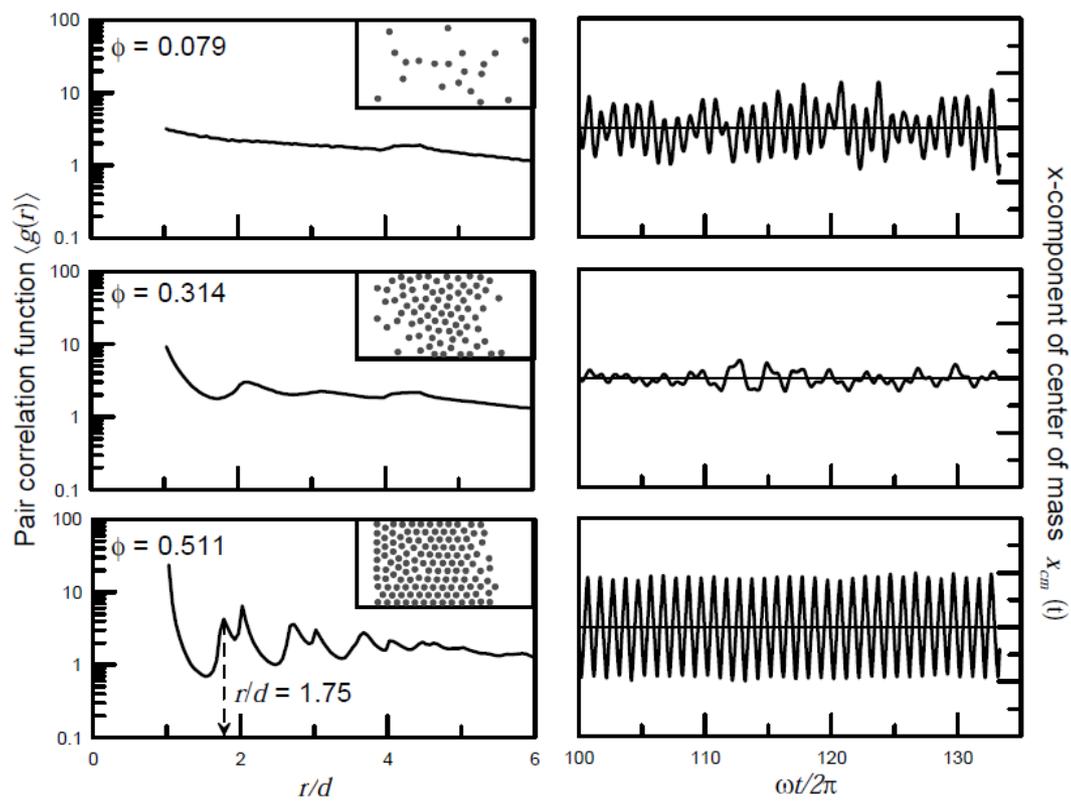

Fig. 5

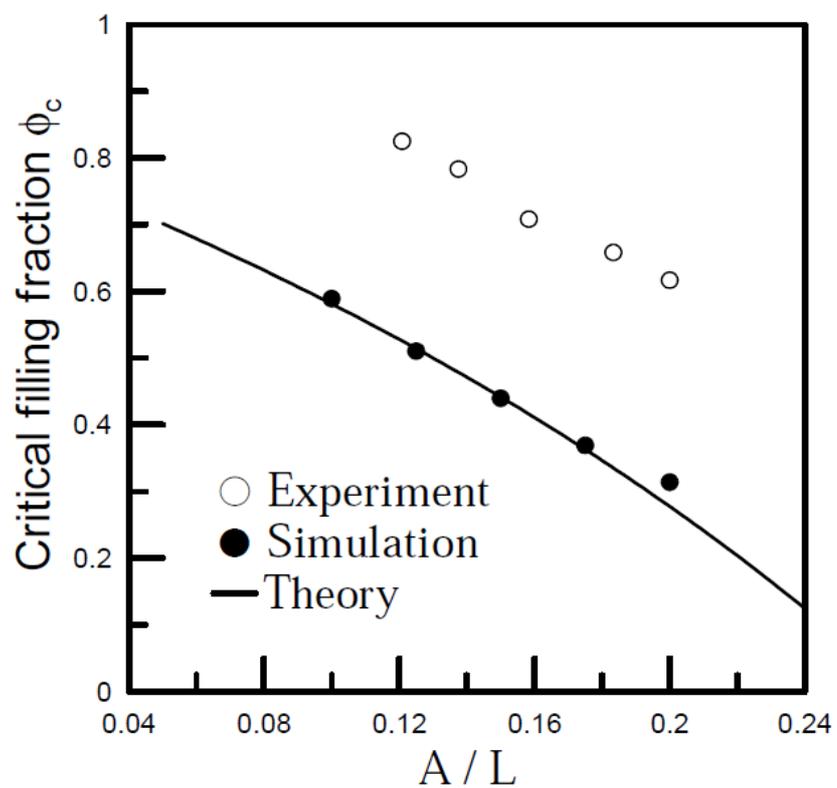

Fig. 6

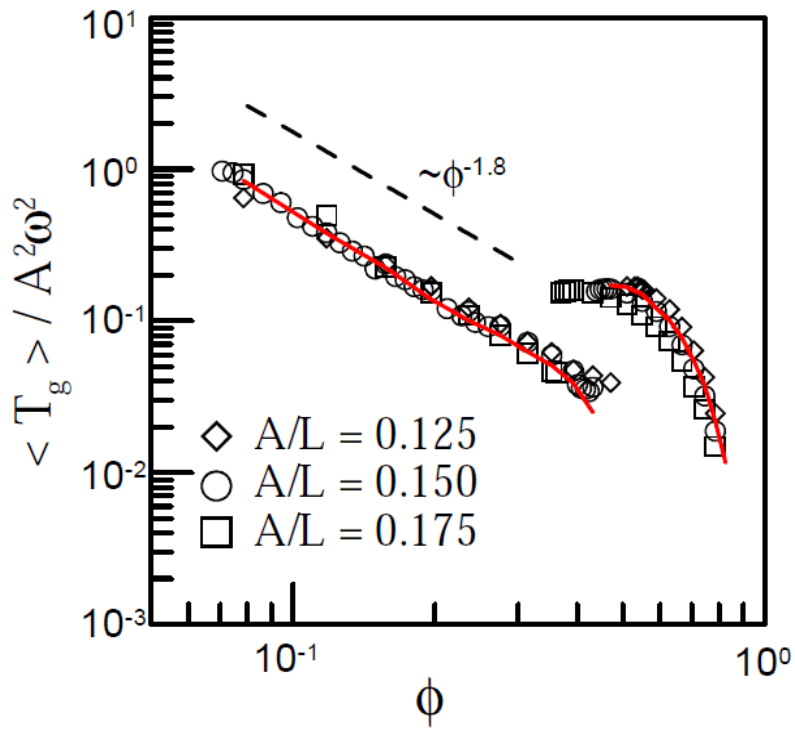

Fig. 7

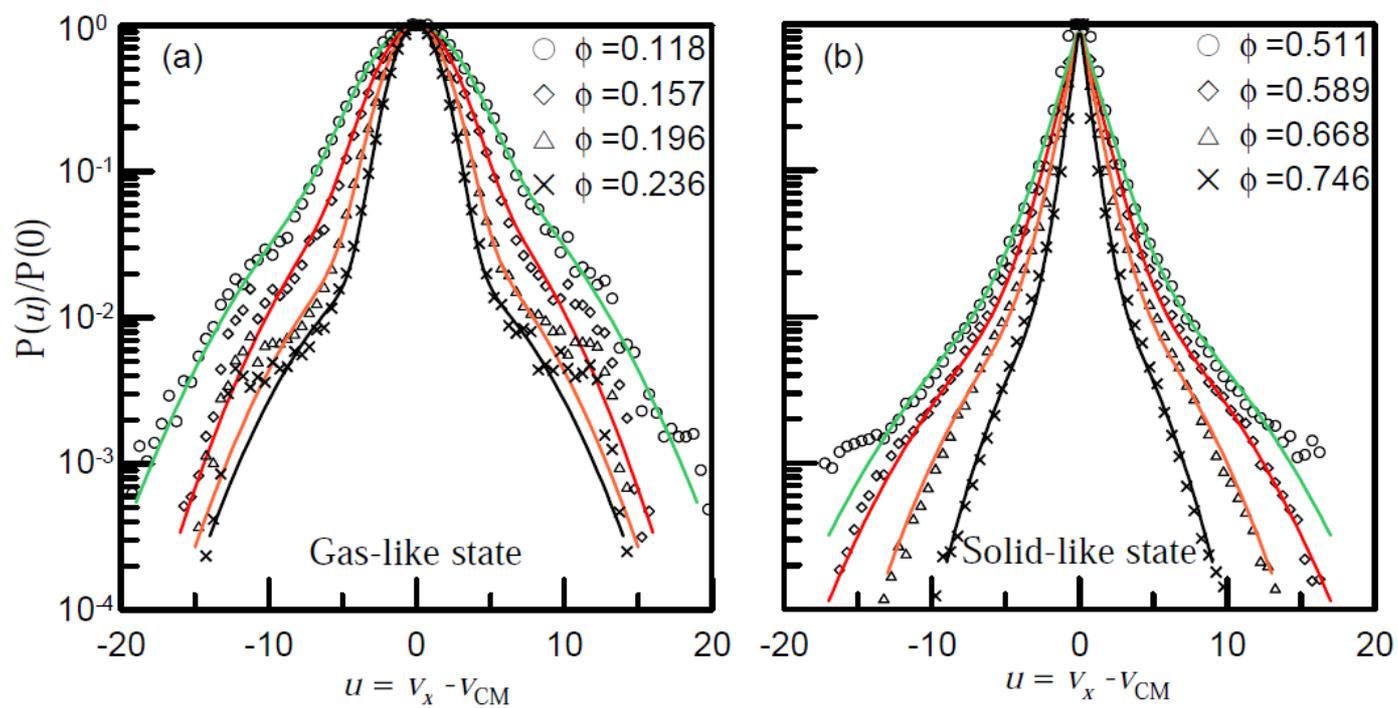

Fig. 8

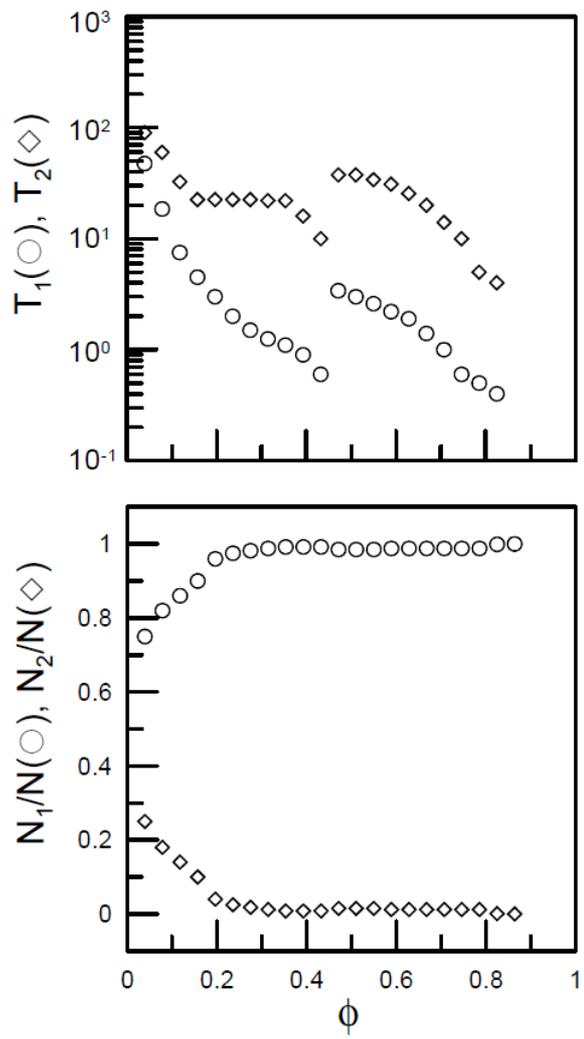

Fig. 9

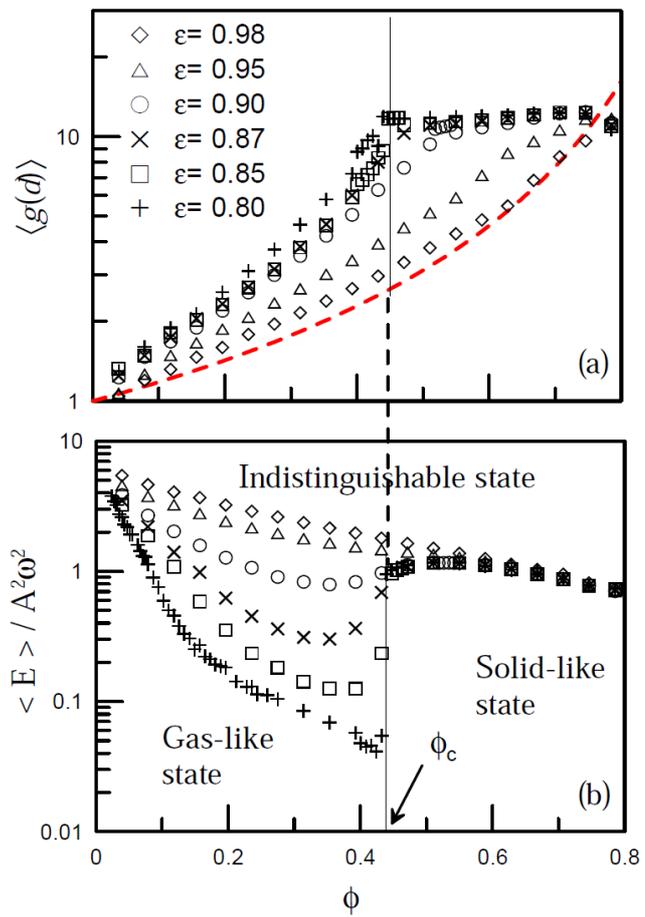